\documentclass[12pt]{iopart}
\usepackage{latexsym}
\usepackage{graphicx}
\usepackage{epsfig}
\usepackage{cite}

\begin{document}

\title[Micro-arcsecond light bending]{Micro-arcsecond light bending by
    Jupiter}\author{M T Crosta\dag\footnote[1]{Present address:
    Observatory of Turin, Strada Osservatorio 20, Pino Torinese (To),
    Italy}and F Mignard\dag }\address{\dag Observatoire de la C{\^o}te
    d'Azur, UMR CNRS 6202, Le Mont Gros, BP 4229, 06304 Nice cedex 4,
    France} \ead{crosta@obs-nice.fr and mignard@obs-nice.fr}

\begin{abstract}
  The detectors designed for Gaia, the next ESA space astrometry
  mission to be launched in 2011, will allow to observe repeatedly
  stars very close to Jupiter's limb. This will open a unique
  opportunity to test General Relativity by performing many
  Eddington-like experiments through the comparison between the
  pattern of a starfield observed with or without Jupiter. We have
  derived the main formulas relevant for the monopole and quadrupole
  light deflection by an oblate planet and developed a simulator to
  investigate the processing of the Gaia astrometric observation in
  the vicinity of the planet. The results show that such an experiment
  carried out with the Gaia data will provide a new fully independent
  determination of $\gamma$ by means of differential astrometric
  measurements and, more importantly, for the first time will evidence
  the bending effect due to the quadrupole moment with a 3$\sigma$
  confidence level. Given the accuracy of the experiment for the
  monopole deflection, this will permit to test alternative modelling
  of the light bending by moving masses.

\end{abstract}

\submitto{\CQG}

\maketitle

\section{Introduction}
\label{section-1}

Several key questions of modern astrophysics regarding the
formation and evolution of the Milky Way will be clarified with
the next space astrometry  mission Gaia, approved in 2000 as a
cornerstone within the European Space Agency science
program~\cite{gaia-book}. The mission is now funded and will enter
in the C/D phase in 2006 for a launch scheduled in late-2011 and
nominal operations over the next five years. Thanks to its high
precision (10 $\mu$arcsecond in angular measurements) and
multi-epoch astrometry, Gaia will be able to detect the relative
positional change of a star resulting for the tiny curvature of
the light ray brought about by the gravitational pull of the Sun,
and to a lesser extent by the giant planets.

As far as the light deflection produced by the solar mass is
concerned, the science case has included from the outset an
investigation of the value of the PPN parameter $\gamma$, considered
as a global unknown in the astrometric model. This parameter, equal
to one in General Relativity (GR), indicates the amount of space-time
curvature produced by a unit rest-mass and it is a fundamental part of
the so called parameterized Post-Newtonian (PPN)
formalism~\cite{will}, originally developed by Eddington for the
famous experiment during the solar eclipse in 1919. The PPN method
aims to quantify the violations of the Equivalence Principle by
parameterizing deviations which modify both local physical laws and
large-scale gravitational phenomena, including the constancy of the
constants. PPN formalism is valid for a broad class of metric theories
and includes GR as a reference case, {\it i.e.} as a current standard
theory of gravitation. The slow motion and weak field limit allows to
use PPN metric expansion as a function of several parameters, varying
from theory to theory, useful to discern gravitational experiments in
the Solar System.  The PPN parameter $\gamma$ is the most important,
since all the other parameters to all relativistic order within the
alternative theories to gravity, converge to their GR value in
proportion to $|1-\gamma|$~\cite{tur}.

The most general formulation of these theories contains an
arbitrary function of a scalar field coupled to the stress-energy
tensor in order to merge quantum mechanics with gravity.  Recent
works on scalar-tensorial theories consistent with various
cosmological scenarios suggest the search for discrepancies from
unity of $\gamma$ at the levels of $10^{-5}$ to $10^{-7}$; more
precisely, Damour and Nordtvedt~\cite{dam-nor} assume that there
exist a cosmological attractor mechanism towards GR quantified
from deviation of $\gamma$ from unity within the above level of
accuracy depending on the total mass density of the Universe.
Then, an experiment purposely designed  could reveal the
coexistence of the two mentioned fields throughout the
cosmological evolution up to the present where pure tensor gravity
explains the gravitational interaction from small to large scale.
In addition, some dynamical models predict scalar fields
dominating the current energy density of the universe, which
should contribute to the value of $\gamma$ at the level of
$10^{-7}$ to $10^{-9}$, thus providing a universe evolution
without the need of dark matter~\cite{tur}. All this issues make
precision tests of gravity in space a fascinating challenge for
the next decades, as in the past the Eddington's observations in
1919 of star line-of-sight did, confirming the amount of
1\hbox{$.\!\!^{\prime\prime}$}72 as deflection angle predicted by
GR.  Nowadays, all present experimental tests are compatible with
the predictions of GR. In particular, experiments conducted in
Solar System have tested all the weak-field predictions of GR
theory at better than the $10^{-3}$ level and down to $2\times
10^{-5}$ level for $\gamma$ as seen from the additional Doppler
shift in radio-wave beams connecting the Earth to the Cassini
spacecraft when they passed near the Sun~\cite{bertotti-gamma}.

New space projects will go deeper in experimental gravitation. The
NASA Gravity Probe B mission~\cite{gpb}, whose first results are
expected very soon, should directly measure $\gamma$ to better
than the $10^{-5}$ level, although this is not the main objective
of the experiment and remains subjected to efficient instrumental
calibrations. An early analysis of the Gaia's capabilities,
indicates that its measurements should provide a precision of
$5\times 10^{-7}$ for this
parameter~\cite{mignard-gamma,ppn-ramod}, an improvement of more
than two orders of magnitude to the current best estimate
mentioned above.  This is quite comparable to the expectations of
the LATOR project~\cite{lator}, which is designed to use two
optical interferometers between two micro-spacecrafts and aims to
determine $|1-\gamma|$ at the level of $10^{-8}$.

Besides the  determination of $\gamma$  based on the global
astrometry capabilities of Gaia, the mission will also allow to
carry out dedicated small field experiments from the observations
performed close to the surface of a giant planet like Jupiter or
Saturn. The light bending due to their gravitational field will be
detectable in the visible and we show in this paper that even the
quadrupole field of Jupiter will be evidenced in such an
experiment. It is therefore possible to carry out the same kind of
eclipse experiment made by Dyson, Eddington, and Davidson in 1920,
but now by comparing stellar positions in the immediate vicinity
of the planets. Not only will this experiment be a further
confirmation of a GR prediction, but it will also help understand
the very difficult question of the light-bending by a moving
source and that of the propagation of gravitation.

Using Jupiter to test GR, Kopeikin and Fomalont claimed they have
determined the speed of gravity trough the VLBI~\footnote{VLBI is the
acronym for Very Long Baseline Interferometry} astrometric measurement
of Shapiro time delay of light from the Quasar J0842+1835, as its
image passed within 3.7 arcmin of the planet (about 10 Jupiter radii, a
far cry from what Gaia can achieve)~\cite{fom-kop}. According to the
post-Minkowskian formalism assumed~\cite{kop}, the gravitational field
due to a moving body propagates with finite speed; consequently, it
influences a photon with some retardation.  The speed of gravity
enters as an extra tiny velocity-dependent corrections to the Shapiro
time delay formula of the order of 4.8 ps. At present there is no
general consensus on the results, especially upon the method to extend
GR to the case where the velocity of gravity differs from the velocity
of light~\cite{carlip,will-c}. In particular, given the observed
limits on $\gamma$, the Fomalont-Kopeikin measurements are not
accurate enough yet to determine the speed of gravity~\cite{carlip}.
Some authors interpret the result within the Lorentz
transformation properties of the weak gravitational field and consider
it as a violation not only of the Lorentz invariance but also of the
Galilean one~\cite{samuel,asada}. Paper~\cite{frittelli} settles the
significance of the Jupiter experiment as a simply standard aberration
of light which propagates across an active medium with an effective
index of refraction induced by the gravitational field of a lens in
motion: the tiny corrections to the Shapiro time delay should be
included into the fitting of microlensing events. However, even if
there is still a controversy about the question of whether the results
depend on the speed of gravity~\cite{kop-cqg} or on the abberation of
light~\cite{asada,frittelli,sanchez}, there is a general agreement
that this measurement is a remarkable precise VLBI detection of a
post-Newtonian effect due to a planet.

The possibility to apply methods of relativistic astrometry to
test GR has  prompted  several groups to model highly accurate
angular observations in the relativistic framework. New formalisms
have been proposed (~\cite{klio, kop, ram-3,lp} and reference
therein) to tackle the relativistic problem of the light path
reconstruction, which, at least at the $\mu$as accuracy, implies
taking into account all the contributions not only due to the bulk
mass of the Solar System bodies, but also to their quadrupole
moment and to the time-dependent gravitational field generated by
their motions (see \tref{table-1}).

\begin{table}
\setlength{\tabcolsep}{8mm}
\begin{center}
  \begin{tabular}{cr@{} lr@{}  lc} \hline \\[-4pt]
     & \multicolumn{2}{c}{$\delta\Phi_\textrm{pN}$} &
       \multicolumn{2}{c}{$\delta\Phi_{Q}$} &
       $\delta\Phi_\textrm{max}$ \\[5mm]
       \hline \\[-4pt]
     Sun      & $1.\!\!''75$ &  & $\sim 1$~$\mu$as  &   &$(180^\circ)$  \\[3mm]
              &   $\mu$as   &  & $\mu$as            &    &              \\
     Mercury  &         $83$ &  &   --              &   &   $9'$ \\
     Venus    &        $493$ &  &   --              &   &   $4.5^\circ $ \\
     Earth    &        $574$ &  &   0.6             &   &$178^\circ$ \\
     Moon     &         $26$ &  &   --              &   &$9^\circ$ \\
     Mars     &        $116$ &  &  0.2              &   &$25'$ \\
     Jupiter  &      $16280$ &  &        239        &   &$90^\circ/3'$ \\
     Saturn   &       $5772$ &  &           94      &   &$18^\circ/51''$ \\
     Uranus   &       $2081$ &  &           25      &   &$72'/6''$ \\
     Neptune  &       $2535$ &  &            9      &  & $51'/3''$ \\
     Pluto    &          $7$ &  &      --           &   &$8''$ \\
    \hline
 \end{tabular}
\caption{\label{table-1} Light deflection amounts
at 1~$\mu$arcsec due to the planets for a photon crossing the Solar
System: pN is for post-Newtonian order and Q for the quadrupolar
moment. The values are computed for grazing rays in the Gaia observing
geometry; figures in the last column give the angular distances
between the perturbing body and the star at which the effect is still
1~$\mu$arcsecond, with Gaia at the Sun-Earth L2. Where two values are
reported they refer, respectively, to pN and quadrupole effect.}
\end{center}
\end{table}

In particular, two papers~\cite{kop-mon} and~\cite{lp-mon} have
gone through the post-Newtonian treatment of the light propagation
in the case of an isolated axisymmetric body, in order to take
into account the influence of the  multipole moments. The former
is based on a mathematical solution of geodesic equation whereby
mathematical expressions for the light deflection in any order of
multipole perturbations are obtained. In the second paper, this
contribution stems from the prior computation of the time transfer
function between two point located at a finite distance. However
neither approach has provided hints on how to apply the intricate
formulas to a practical case like the one faced in the Gaia
modelling.

Having in mind the capabilities of Gaia to observe stellar sources
very close to Jupiter's limb (at a fraction of the planet radius),
we have evaluated the light deflection produced by an oblate
planet on grazing photons coming from distant stars. This study is
part of a  wider project called GAia Relativistic EXperiment
(GAREX), which aims to investigate all possibilities to test GR
with Gaia measurements. In the solar system, this mission will
mainly carry out:
\begin{itemize}
    \item light deflection experiments, divided into (i) global
      astrometry, in particular highly accurate determinations of
      $\gamma$  by observing the change in  star position at different angular distance to the Sun;
       (ii) small field experiments, with the examination of the
       light propagation by  means of differential
      measurements of stellar positions near the planets;
\item perihelion precession effect, related to the determinations
of
  PPN parameter $\beta$ \footnote{The PPN parameter $\beta$ measures the
  non-linearity in the superposition of gravitational fields.}
  from the orbit fitting of several thousands  minor bodies~\cite{hestro}.
 \end{itemize}
 In this paper we concentrate on the case of the light bending by Jupiter at the
 microarcsecond level and its detection by the small field astrometric observations with Gaia.
  In~\cite{poster-paris} we have already  explored the favorable circumstances for Gaia in five years of
 continuous observations, approximately from 2012 to 2017, to detect
 the quadrupole light bending effect predicted by GR, but never
 observed.  For Jupiter the magnitude of the monopole deflection for a
 grazing ray is $\sim 16$ milliarcsecond, to which a component from
 the quadrupole moment is superimposed with an amplitude of $\sim 240
 $ micro-arcsecond, as showed in
 \tref{table-1}. This secondary deflection has a
 very specific pattern as a function of (i) the position of the star
 with respect to the oblate deflector and (ii) the orientation of its
 spin axis.  In \sref{section-2} we derive the relevant
 formulas to express this light bending effect using the PPN formalism
 at large and small angle from the planet. In \sref{section-3}
 we describe all the steps needed to simulate the small field
 experiment when observing stellar background in vicinity of
 Jupiter. Finally, in \sref{section-4}, we discuss the data
 processing of a set of Montecarlo runs and the results in the
 determination of the PPN Parameter $\gamma$ and the quadrupole effect
 from Jupiter.

\section{The light deflection produced by an axisymmetric planet}
\label{section-2}

Most stellar sources, planets and observers have a small velocity
compared to the velocity of the light and a weak inter-body
gravitational field exists inside bound systems. Moreover, if we
consider isolated distribution of matter, the geometry can be
assumed Minkowskian, asymptotically flat far away~\cite{misner}. Let us
consider, then, a set of PPN coordinates $\{x^{\alpha}\}$
($\alpha=0,1,2,3$) and a solar system locally perturbed by
isolated stationary axisymmetric masses. To the post-Newtonian
(pN) accuracy, the spatial part of the geodesic equation of a
light ray coming form a distant star can be easily transformed into~\cite{will}:
\begin{equation}\label{ppn-i-geodetic}
\frac{d^2 x^i}{d {x^{0}}^2} =  U_{,i} \left(1 + \gamma \,
\delta_{jk} \frac{d x^j}{d x^0} \frac{dx^k}{d x^0}\right)- 2
(1+\gamma) \frac{dx^i} {dx^0} \left( \delta_{jk} \frac{dx^j}{d
x^0} U_{,k} \right),
 \end{equation}
where $U= M/\sqrt{(\delta_{ij} x^i x^j)}$ ($i,j,k=1,2,3 $) is the
Newtonian potential (in geometrized units) and $ U_{,i}$ its
partial derivative with respect to the spatial coordinates. A
solution of the above equation can be get by tracing a Newtonian
zero order straight-line $x^i_N$ plus all the relativistic
deviations $x^i_D$, \textit{i.e.}
\begin{equation}\label{stright-line}
 x^i = x^i_N + x^i_D.
 \end{equation}
 Then, by decomposing $x^i_D$ into two components,
 parallel ($x^i_{D_{\parallel }}$) and perpendicular ($x^i_{D\bot }$) to
 $x^i_N$, equation \eref{ppn-i-geodetic} becomes
\begin{equation}
\label{ppn-xD-1} \frac{d^2 x^i_{D \bot}}{d{x^{0}}^2} = (1+
\gamma)[ U,_{i} - t^i (\delta_{jk} t^j U,_{k})]=
(1+\gamma) \mathbf{\nabla}_{\bot}U,
\end{equation}
where $ \mathbf{t} = dx^i/dl$ is the unit tangent vector on the
unperturbed light path in the direction of the observer. Here the
symbol $\mathbf{\nabla}_{\bot}$ indicates the derivative
perpendicular to the light ray, assumed to pass outside the matter
distribution.  Placing the origin of coordinates $x^i_0$ at the
center of an axisymmetric planet, \textit{i.e.} $\mathbf{z}$ considered as
its axis of maximum moment of inertia, located between the star
and the observer, the positional vector of the photon with respect
to the principal axes centered at the planet can be expressed as:
\begin{equation}
r^i =  x^i - x^i_{0} = t^i \ell - n^i b,
\end{equation}
where $\ell$ represents the length of the light path and $\mathbf{n}$ the
radial direction perpendicular to the unperturbed ray pointing towards the
center of gravity, along the impact parameter $b$.

\begin{figure}[htp]
  \begin{center}

\includegraphics[scale=0.5, clip=true]{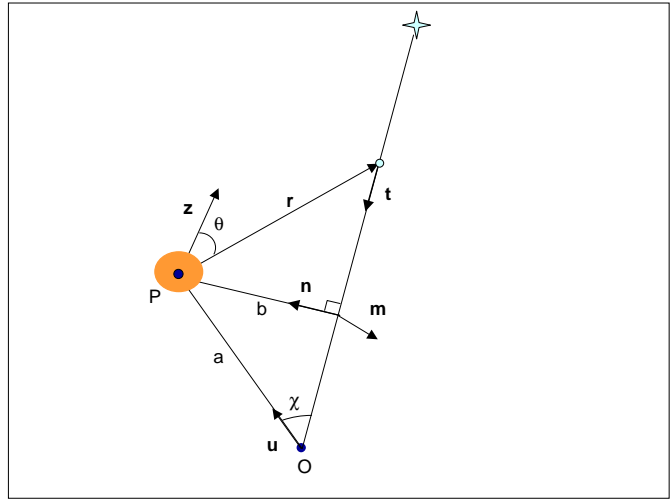}
   \end{center}
  \caption{\label{figure-1}Geometry of light deflection due to a
  planet (P): the spin axis of the planet \textbf{z} is out of the
  plane; \textbf{t} represents the unit tangent vector from a
  distant star (S) to the observer (O) on the unperturbed light
  trajectory; \textbf{u} is the unit direction from O to P along their
  distance a; finally, $\chi$ is the angle S\^OP, and b the impact
  parameter.}
\end{figure}

\begin{figure}[htp]
  \begin{center}
\includegraphics[scale=0.5,clip=true]{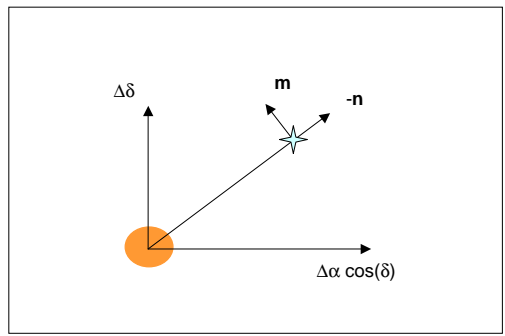}
   \end{center}
  \caption{\label{figure-2}Light deflection by a planet:
    tangent plane on the sky.  The position of the star is displaced both in
    the radial (-$\mathbf{n}$) and orthoradial $\mathbf{m}$
    directions. The spin axis of the planet $\mathbf{z}$ (not shown here) does not
    lie  in this plane in general.}
\end{figure}

The light deflection is given by the deviation of the tangent
vector $\mathbf{t}$ from the straight-line along the photon's path
as,
\begin{equation}
\label{deflected-vector}
\Delta \mathbf{\Phi} \equiv
\int \frac{d \mathbf {t}}{d \ell} d \ell
\end{equation}
which leads with \eref{ppn-xD-1} to,
\begin{equation}
\label{dphi} \Delta \mathbf{\Phi} =(1+\gamma)\int
\mathbf{\nabla}_{\bot}U\, dl.
\end{equation}
 By taking the gravitational potential up to the quadrupole
term one gets,
\begin{eqnarray}
\mathbf{\nabla}_{\bot} U &=& \left[- \frac{b}{r}
\left(-\frac{M}{r^{2}} + \frac{3M}{r^{4}} J_{2} R^{2} \frac{5
\cos^{2} \theta - 1}{2} \right) \right. \nonumber \\ && \left. +
\left(-\frac{3 M}{r^{4}} J_{2} R^{2} \cos\theta\right) (\mathbf{z}
\cdot \mathbf{n} )\right] \mathbf{n}\label{grad_p-U} \\ &&
+\left(-\frac{3 M}{r^{4}} J_{2} R^{2} \cos\theta\right)(\mathbf{z}
\cdot \mathbf{m} )\mathbf{m} \nonumber
\end{eqnarray}
where $r$ and $\theta$ are respectively the radial distance and
the co-latitude (measured from the planet north pole) of a field
point on the light trajectory, $J_{2}$ is the dimensionless
coefficient of the second zonal harmonic, $R$ the radius and $M$
the mass of the planet, and, finally, $\mathbf{m}$ represents the
orthoradial component (see \fref{figure-1} and \fref{figure-2}).

The length $\ell$ of the photon path can be scaled by the impact
parameter into a dimensionless parameter as follows
\begin{equation} d\ell= b d\lambda, \label{lambda-param}
\end{equation}
so the radial distance becomes
\begin{equation}
r= b(1 + \lambda^{2})^{1/2}. \label{r-param}
\end{equation}
Each integral entering~\eref{dphi} must be computed with
$\lambda$ running positively in the same direction as the photon
from $\lambda=-\infty$ to $\lambda= 1/\tan\chi$, with $\chi$
standing for the angular separation between the directions
star/observer and observer/planet (\fref{figure-1}).
At the closest approach on the unperturbed ray one has
$\lambda=0$. The explicit expressions of  the integrals are given
in appendix A. After some algebra, the light deflection vector is
split into two components, the first one along $\mathbf{n}$ and
the second one along $\mathbf{m }$, both including the monopole
and the quadrupole contribution of the planet in function of the
angular separation $\chi$:
\begin{equation}
\Delta \mathbf{\Phi}= \Delta \Phi_{1}\mathbf{n} + \Delta\Phi_{2}\mathbf{m},
\label{deftot}
\end{equation}
where, precisely,
\begin{eqnarray}
\Delta\Phi_{1}&= (1+\gamma) \frac{2 M}{b} \left
\{(1+\cos\chi)+J_{2}\frac{R^{2}}{b^{2}} \left
[(1+\cos\chi+\frac{1}{2}\cos\chi\sin^{2}\chi)
\right. \right. \nonumber \\ &
\left. \left. -2(1+\cos\chi+\frac{1}{2}\cos\chi\sin^{2}\chi+
\frac{3}{4} \cos\chi\sin^{4}\chi)(\mathbf{n}\cdot \mathbf{z})^{2}
\right.\right. \nonumber \\ &
\left.\left.+(\sin^{3}\chi-3\sin^{5}\chi)(\mathbf{n}\cdot\mathbf{z})(\mathbf{t}\cdot\mathbf{z})\right.\right.
\label{defqu1-chi}\\
&-\left.\left.(1+\cos\chi+\frac{1}{2}\cos\chi\sin^{2}\chi-\frac{3}{2}\cos\chi\sin^{4}\chi)(\mathbf{t}\cdot\mathbf{z})^{2}
\right]\right\} \nonumber
\end{eqnarray}
and
\begin{eqnarray}
\Delta\Phi_{2}&= \frac{(1+\gamma) MJ_{2}R^{2}}{b^{3}}
\left[2(1+\cos\chi
+\frac{1}{2}\cos\chi\sin^{2}\chi)(\mathbf{n}\cdot\mathbf{z})
(\mathbf{m}\cdot\mathbf{z}) \right. \nonumber\\
&\left.+\sin^{3}\chi(\mathbf{m}\cdot\mathbf{z})(\mathbf{t}\cdot\mathbf{z})\right].
\label{defqu2-chi}
\end{eqnarray}
The first term in the radial component (that along $\mathbf{n}$)
is the classical monopole deflection, widely used in astronomy.
All the other terms factored by $J_{2}$ come from the quadrupole
of the planet. One can specialize these expressions to
near-grazing rays as, unlike with the solar deflection, the effect
is too small to be observed at large angle from the planet. Hence
when $\chi \ll 1$ this leads to the more convenient and accurate
enough formulas,
\begin{equation}
\Delta\Phi_{1} = \frac{2 (1+\gamma) M}{b}\left[1+
J_{2}\frac{R^{2}}{b^{2}} \left(1-2(\mathbf{n}\cdot\mathbf{z})^{2}
-(\mathbf{t}\cdot\mathbf{z})^{2} \right) \right] \label{defqu1}
\end{equation}
\begin{equation}
\Delta\Phi_{2}=\frac{4(1+\gamma) M J_{2}
R^{2}}{b^{3}}(\mathbf{m}\cdot\mathbf{z})
(\mathbf{n}\cdot\mathbf{z}). \label{defqu2}
\end{equation}
which are similar to a derivation obtained in~\cite{shapiro} in
the case of the Sun.

The deflection vector depends on the orientation of the spin axis
of the planet and, moreover, on the direction of the star with
respect to the planet as seen in the terms proportional to
$(\mathbf{n}\cdot\mathbf{z})$ and $(\mathbf{t}\cdot\mathbf{z})$.
In general the quadrupole deflection vector depends both on the
impact parameter and on the direction of the light source with
respect to the spin axis projected on the plane of the sky.
However, for a given impact parameter, its modulus \eref{deftot}
depends only on the star's direction with respect to the planet's
reference frame. This property is established in Appendix B and
has been used thoroughly to as an additional check to the computer
implementation of the simulation (see \sref{section-3.2}). One
should note that the above formulation has left aside and
neglected refinements in the modelling like the motion of Jupiter,
the retarded effect on the light propagation and mulipolar terms
of higher orders in the potential expansion.

 Considering $\gamma=1 $ is equivalent to saying that
equations~\eref{defqu1}-\eref{defqu2} constitute the reference
modelling in the standard GR. Therefore, in the same way as one
introduces $\gamma$ in the monopole deflection to characterize
departures from GR, we multiply $J_2$ by a dimensionless parameter
$\epsilon$ whose value is 0 if no effect is seen (say this may
correspond to a lack of coupling between gravity and
electromagnetism seen in the multipole moments) and 1 for the
standard GR. Hereafter we call this parameter the Quadrupole
Efficiency Factor (QEF) and the goal is to see whether it is
significantly determined from the Gaia observations. Although
$\epsilon$ is not a PPN parameters, it represents the first level
of a post-Newtonian test whose goal is to determine whether the
effect is detected,\textit{i.e.} if one can say that $\epsilon$ is
equal to unity, with a certain uncertainty. Thus, for the
processing  of the experiment of light deflection near the planet,
we adopt the following formulation,
\begin{equation}
\Delta \Phi_{1} = \frac{2(1+\gamma) M }{ b}\left[1+ \epsilon J_2
\frac{R^{2}}{b^{2}} \left(1-2(\mathbf{n}\cdot\mathbf{z})^{2}
-(\mathbf{t}\cdot\mathbf{z})^{2} \right) \right] \label{defqu1-gam-eps}
\end{equation}

\begin{equation}
\Delta \Phi_{2}=\frac{4M (1+\gamma) \epsilon {J_2} R^{2}
\gamma}{b^{3}}(\mathbf{m}\cdot\mathbf{z}) (\mathbf{n}\cdot\mathbf{z}).
\label{defqu2-gam-eps}
\end{equation}
including the two unknowns $\gamma$ and $\epsilon$, which are
equal to $1$ in the simulation.

\section{Small field experiments with Jupiter}
\label{section-3}

The one GAREX experiment we consider in this paper is a light
deflection  in the vicinity of Jupiter. The goal is to measure in
a fully independent way the two parameters $\gamma$ and $\epsilon$
expressed in light deflection formulae~\eref{defqu1-gam-eps}
and~\eref{defqu2-gam-eps}, together with their associated errors,
knowing the astrometric accuracy that will be achieved by Gaia. As
$J_2$ is well known from space probe tracking with a precision
better than $10^{-5}$, we will here consider it as a constant, and
infer instead an estimate of the QEF factor $\epsilon$.

\subsection{Ephemerides}
\label{section-3.1} 
\begin{figure}[h]
 \begin{center}
   \leavevmode
 \centerline{\epsfig{file=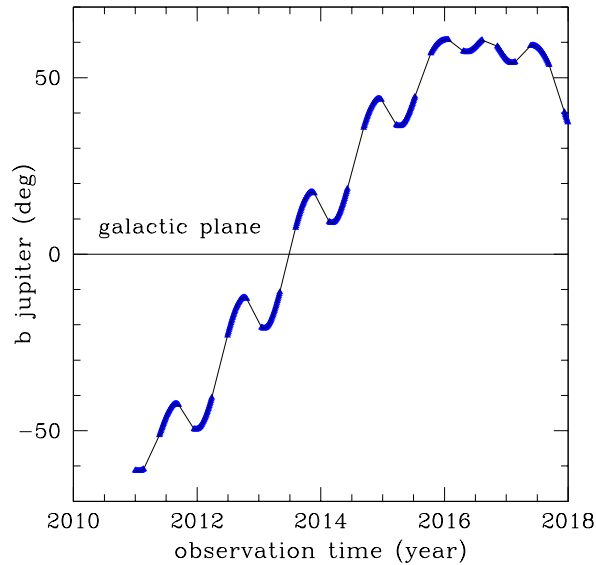, width=0.5\linewidth,clip=true}}
\end{center}
 \caption{\label{figure-3}Galactic latitude (b) of Jupiter versus the observing
   time (in years). The solid line represents the galactic latitude of Jupiter
   (as seen from Gaia) over the years 2011-2018 and the highlighted patches
   correspond to the visibility periods when the angular distance to the Sun
   makes the observation possible.}
  \end{figure}

First, we have investigated what are the
favorable circumstances to perform experiments using Jupiter
during the mission lifetime~\cite{poster-paris}. This depends on
the number of times Jupiter will cross one of the astrometric
fields during the mission and on the stellar density in its
immediate surroundings during these observations. At present, the
details of the former can only be known statistically, as the
precise initial conditions of the sky scanning law remain unknown.
On the other hand the sky background can be assessed with more
certainty as it depends only on the motion of Jupiter between 2012
and 2018, and not very accurately on when the observations take
place, provided they are reasonably scattered during the observing
period. The ephemerides of Jupiter have been computed in galactic
coordinates between these two epochs for an observer orbiting the
Sun-Earth L2 in the same way as Gaia will do. The observability
conditions are shown in~\fref{figure-3} with the visibility
periods when the angular distance to the Sun makes the
observations feasible with Gaia (no observations can be made
around conjunctions and oppositions). Galactic coordinates are
used just because the primary dependence of the stellar density is
on the galactic latitude.
\begin{figure}[htp]
 \begin{center}
 \includegraphics[scale=0.4, clip=true]{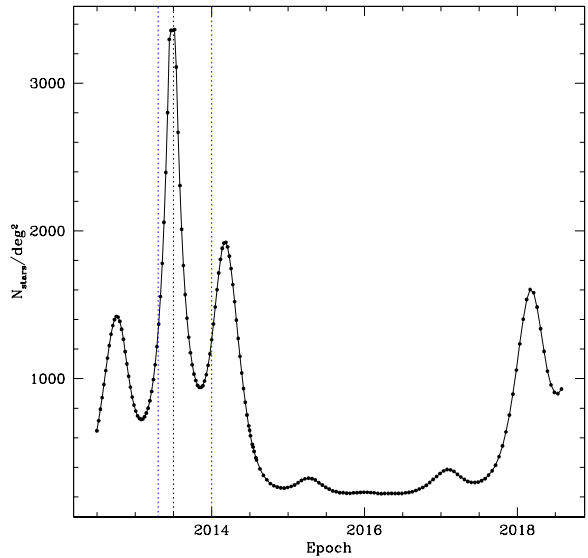}
\end{center}
 \caption{ \label{figure-4} Mean number of stars per square
 degree down to the magnitude limit V=20 in the field of Jupiter
 during the Gaia operations. The dashed lines represent the period when
 Jupiter crosses the galactic plane and the central one the epoch
 when the maximum number of stars is observed.}
\end{figure}

We have chosen the date 2012 as the beginning of the simulations,
since Gaia will take at least six months to be operational at the
Lagrangian point L2 of the Earth-Moon-Sun system. Jupiter will
cross the galactic plane in mid-2013 and, consequently, we expect
favorable observations of the planet in front of a dense stellar
background as shown in~\fref{figure-4}. The plot
gives the average number of stars per square degree around Jupiter
from 2012 to 2018 as seen from Gaia.

\subsection{Simulation of the experiment}
\label{section-3.2}

The principle of the simulated experiment is to generate a series
of pairs of observations over the mission length of 5 years, so
that in each pair Jupiter is, or is not, at the center of the
stellar field. In both cases astrometric observations similar to
that expected from Gaia are computer generated, with the proper
noise function of the star brightness. The number of stars
simulated in each field is determined by the galactic latitude of
Jupiter at the observation time. Then the two star patterns are
compared and the differences in along-scan positions (the only
direction of accurate measurements with Gaia) are fitted to the
model with the two unknowns $\gamma$ and $\epsilon$. This is an
idealized version of the Gaia procedure as the two fields (with or
without Jupiter) will not be observed exactly a the same time,
leading to correction for the relative orientations of the fields
and for the proper motion of the stars. But these are technical
details that will be carefully handled during the data processing
and have no impact on the feasibility study demonstrated in this
paper. The numerical parameters used in the simulation are listed
in \tref{table-2}.
\begin{table}[htp]
  \caption{Main parameters used for the simulation.}
  \label{table-2}
  \begin{center}
    \leavevmode
        \begin{tabular}[h]{lrcc}
      \hline \\ [-5pt]
    parameter &  numerical value \\
   \hline \\ [-5pt]
    number of observations & 90  \\
    area of the field-of-view & 0.6 deg $\times$ 0.6 deg \\
    magnitude range &  12.0 to 20.0 \\
    $R_J$  & $7.13 \times 10^7$ m\\
    $M = GM_J/c^2 $  &        1.41 m   \\
    Jupiter $J_2$   & 0.014736 \\
    $\gamma$ & 1 \\
    $\epsilon$ &   1\\
    $\sigma_{\mbox{pos}}$  along the scan at $V=15$        & 100$\mu$as\\
    $\sigma_{\mbox{pos}}$  across the scan at $V=15$       & 300$\mu$as \\
    \hline
    \end{tabular}
  \end{center}
\end{table}

\begin{figure}[htp]
 \begin{center}
\includegraphics[scale=0.4,clip=true]{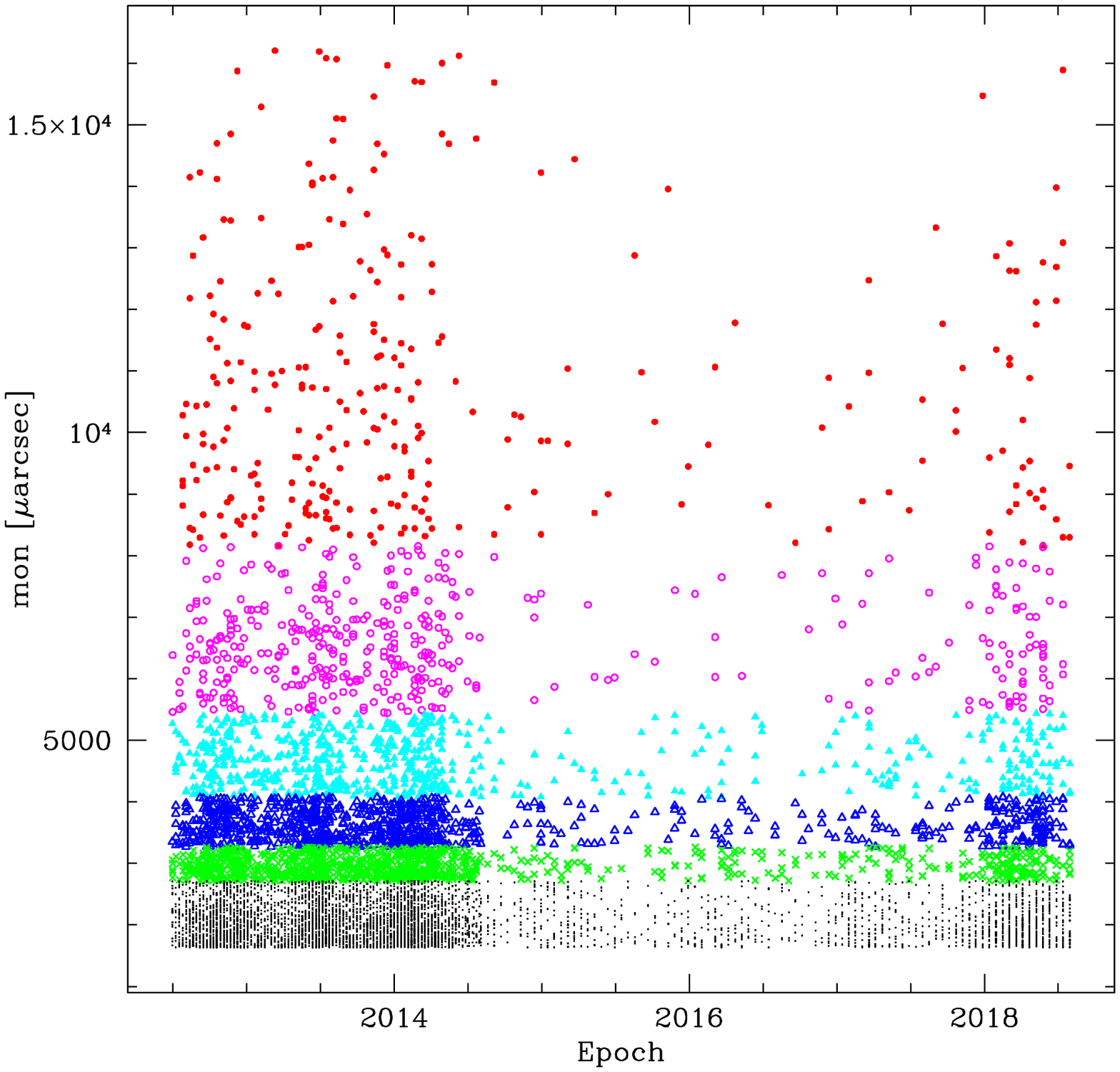}\includegraphics[scale=0.4,clip=true]{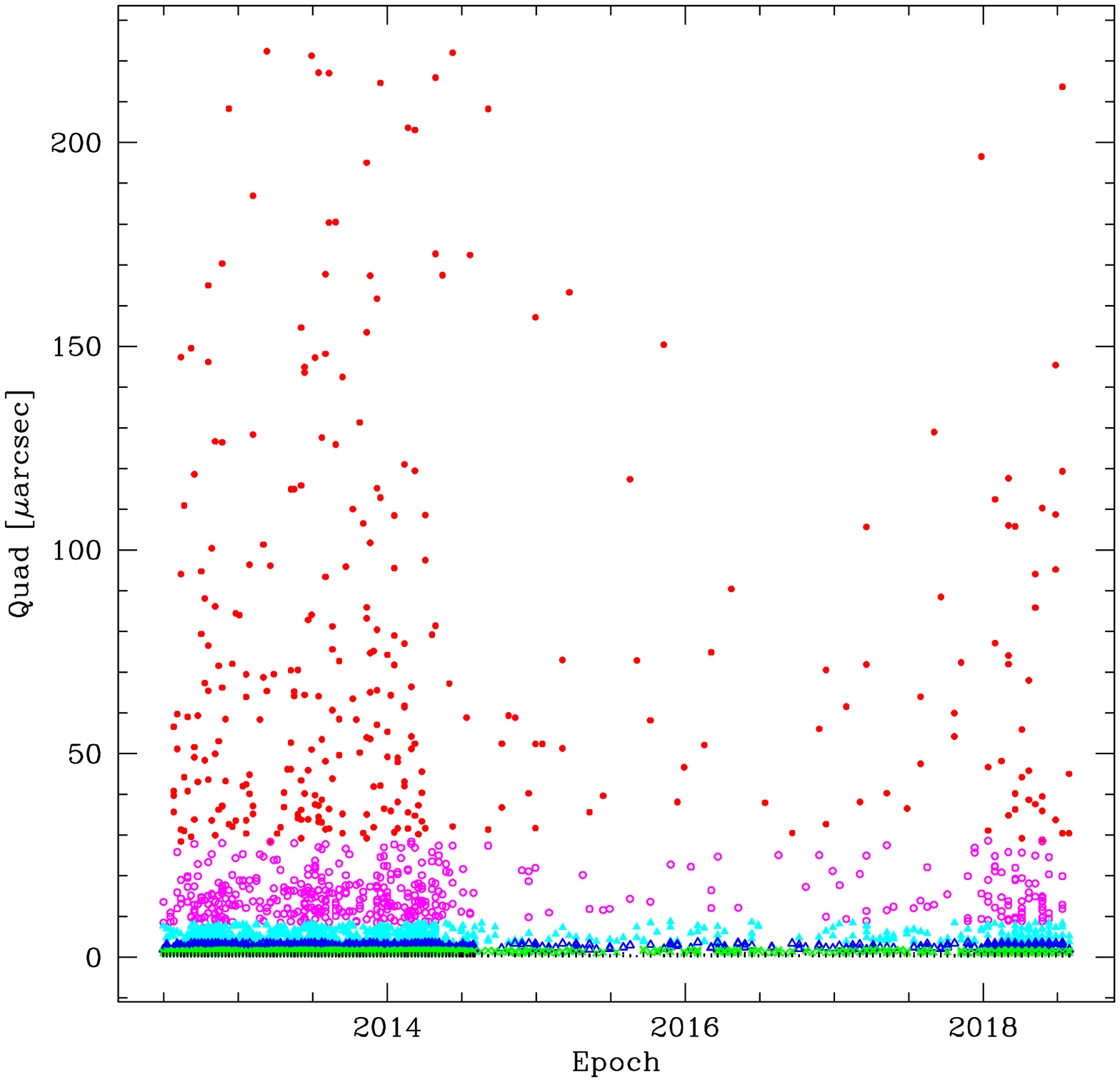}
\caption{\label{figure-5-6}
The monopole (left-hand side) and quadrupole (right-hand side)
deflections over all epochs of Jupiter's visibility for all stars up
the magnitude limit V=20, where $1 < b/R_J \le 2$ (full square),
$2 < b/R_J  \le 3$ (open square), $3< b/R_J \le 4$ (open triangle),
$4<b/R_J \le 5$ (full triangle), $5< b/R_J \le 6$ (cross), $ b/R_J > 6$ (dots).} \end{center}
\end{figure}

\begin{figure}[htp]
\begin{center}
\includegraphics[scale=0.4]{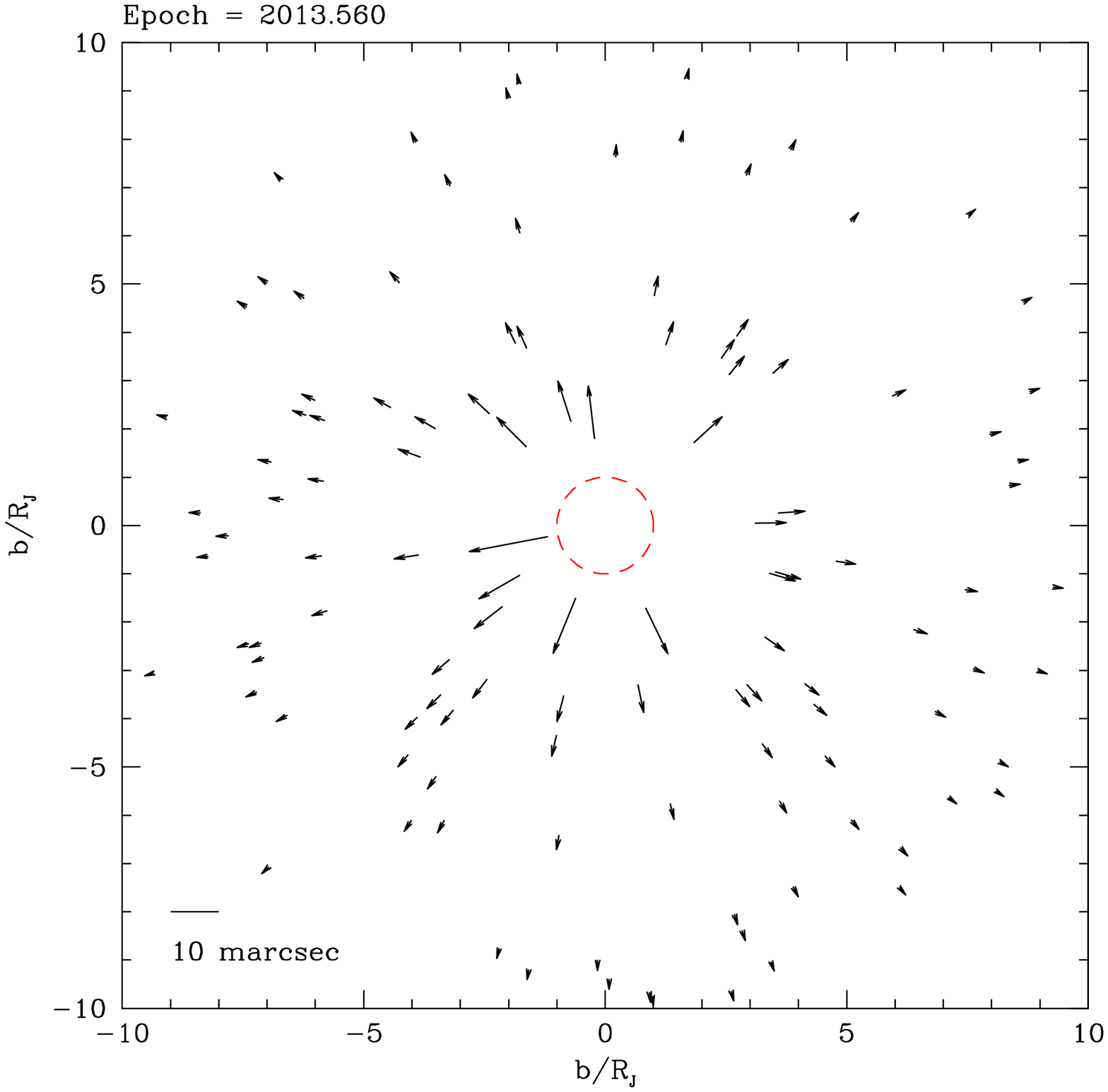}\includegraphics[scale=0.4]{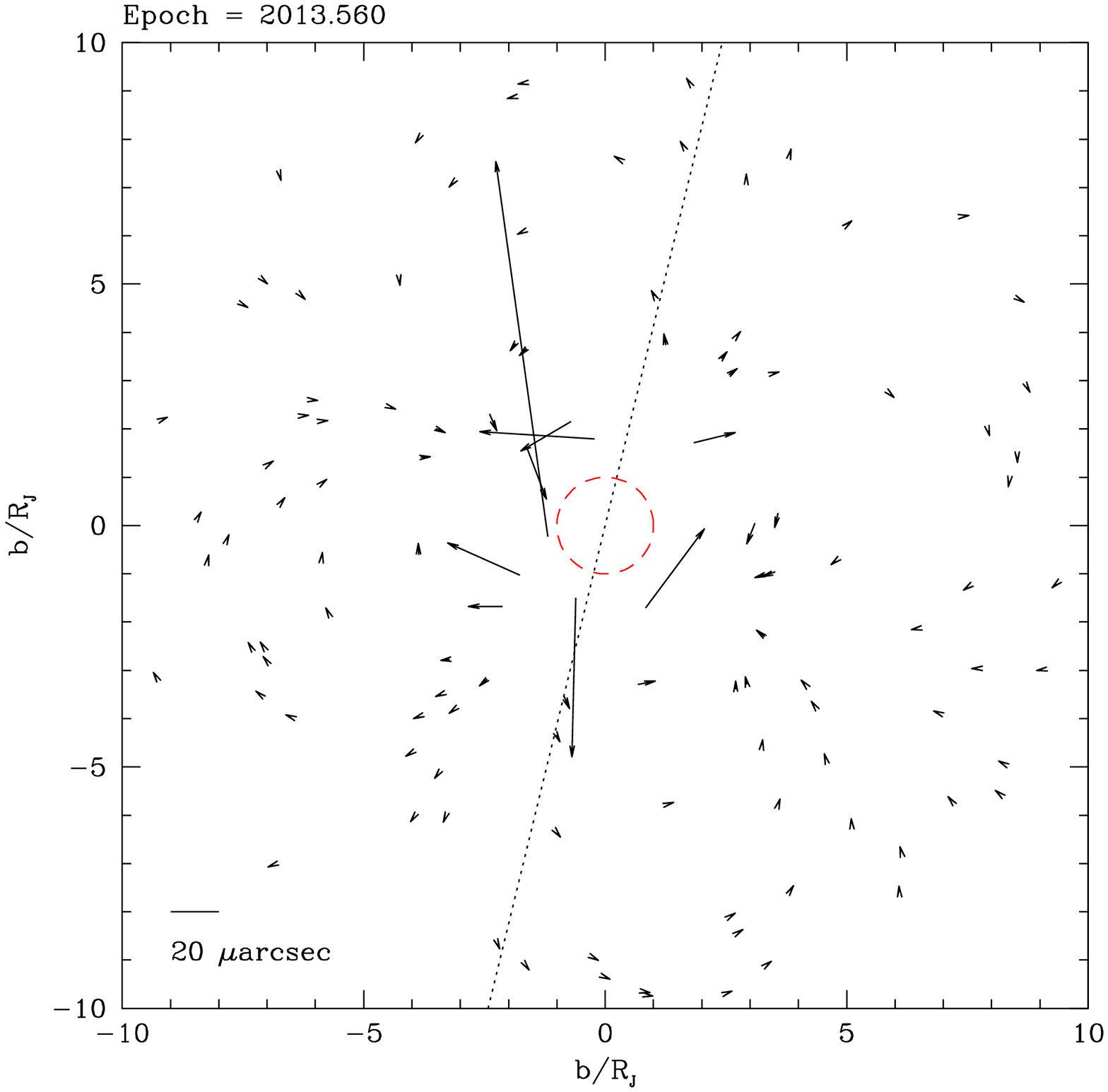}
\caption{\label{figure-7-8}
Observer view of the monopole (left-hand side) and quadrupole
(right-hand side) light deflection vector field around Jupiter
(circle) in mid 2013. The scales of 10 marcsecond for the monopole and 20
$\mu$arcsecond for the quadrupole are shown on the lower left of the plots,
while the dotted line indicates the direction of the Jupiter's spin
axis with respect to the line-of-sight. } \end{center}
\end{figure}

In order to use
equations~\eref{defqu1-gam-eps}-\eref{defqu2-gam-eps} in the
simulation, we have fixed the origin of the coordinate system at
Jupiter's center and assumed that it coincides with the center of
the astrometric focal plane. Given the galactic coordinates
$(l,b)$ of Jupiter we extracted the relevant stellar density in
the magnitude range $V=12.0-20.0$ from a realistic galactic model
based in actual star counts fitted to convenient mathematical
expressions as a function of the galactic latitude and longitude.
The magnitude range is determined by the rejection of saturated
stars (brighter than V=12.0) on the bright side and by keeping
stars observable with Gaia with a reasonable astrometric accuracy
on the faint side.  One must keep in mind that the number of close
approaches between Jupiter and stars brighter than $V \simeq 12$
will be a rare event, albeit very important in this context, but
not liable to statistical treatment. It must be handled on a case
by case basis with a real sky and this will be done later when the
final parameters of the scanning law are known. After having
computed the number of stars per interval of 0.5 magnitude inside
the field of view at each observational epoch, we generated the
observed stars in the field of view. For the fields with Jupiter
we have computed the proper  stellar directions by taking into
account the full deflection due to the monopole and the
quadrupole. Then for any pair and any epoch we have added a
gaussian noise in both coordinates. The standard deviation of the
noise is determined by the magnitude of the star and the
coordinates (along- or across-scan) with~\cite{gaia-book}:
\begin{equation}
\label{accuracy}
\sigma_{als} = a\times 10^{0.2 (V -15)}
\end{equation}
for stars with magnitude $V > 12.5$. The coefficient $a$ is the
scaling factor giving the single observation astrometric accuracy
for a star of 15 magnitude. For stars brighter than 12.5,
$\sigma_{als}$ is constant and chosen to be 30~$\mu$as for the
measurements along the scan and 100~$\mu$as across.  We have used
the property expressed in \eref{B-quad-mod} to cross-check the
contribution of the quadrupole term in the simulation.  Other
checks were also performed in different reference frames that
helped improve the reliability of the simulator. \Fref{figure-5-6}
and \fref{figure-7-8} illustrate several important features of the
simulated data.
\begin{figure}[htp]
  \begin{center}
\includegraphics[scale=0.4,clip=true]{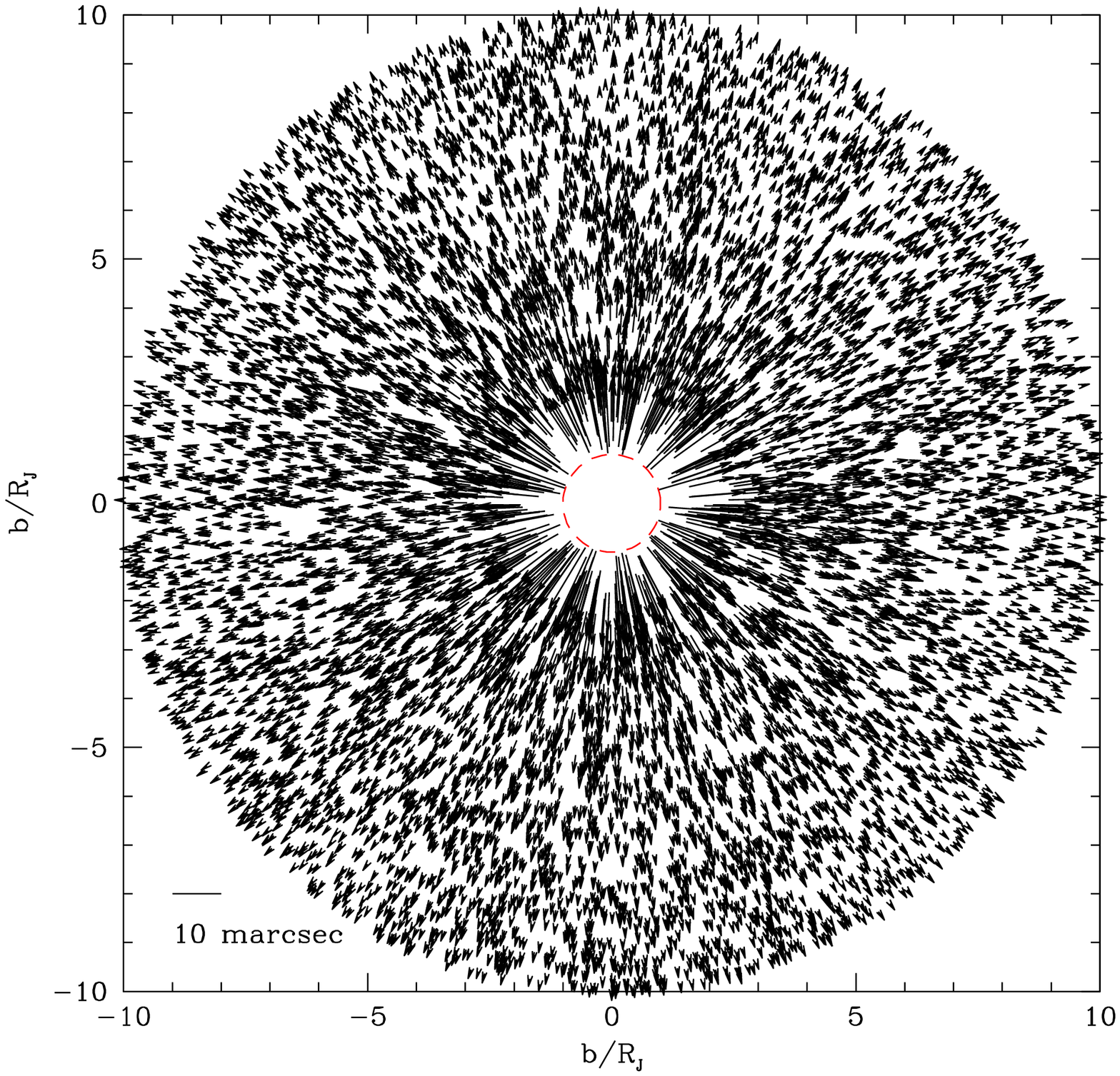}\includegraphics[scale=0.4,clip=true]{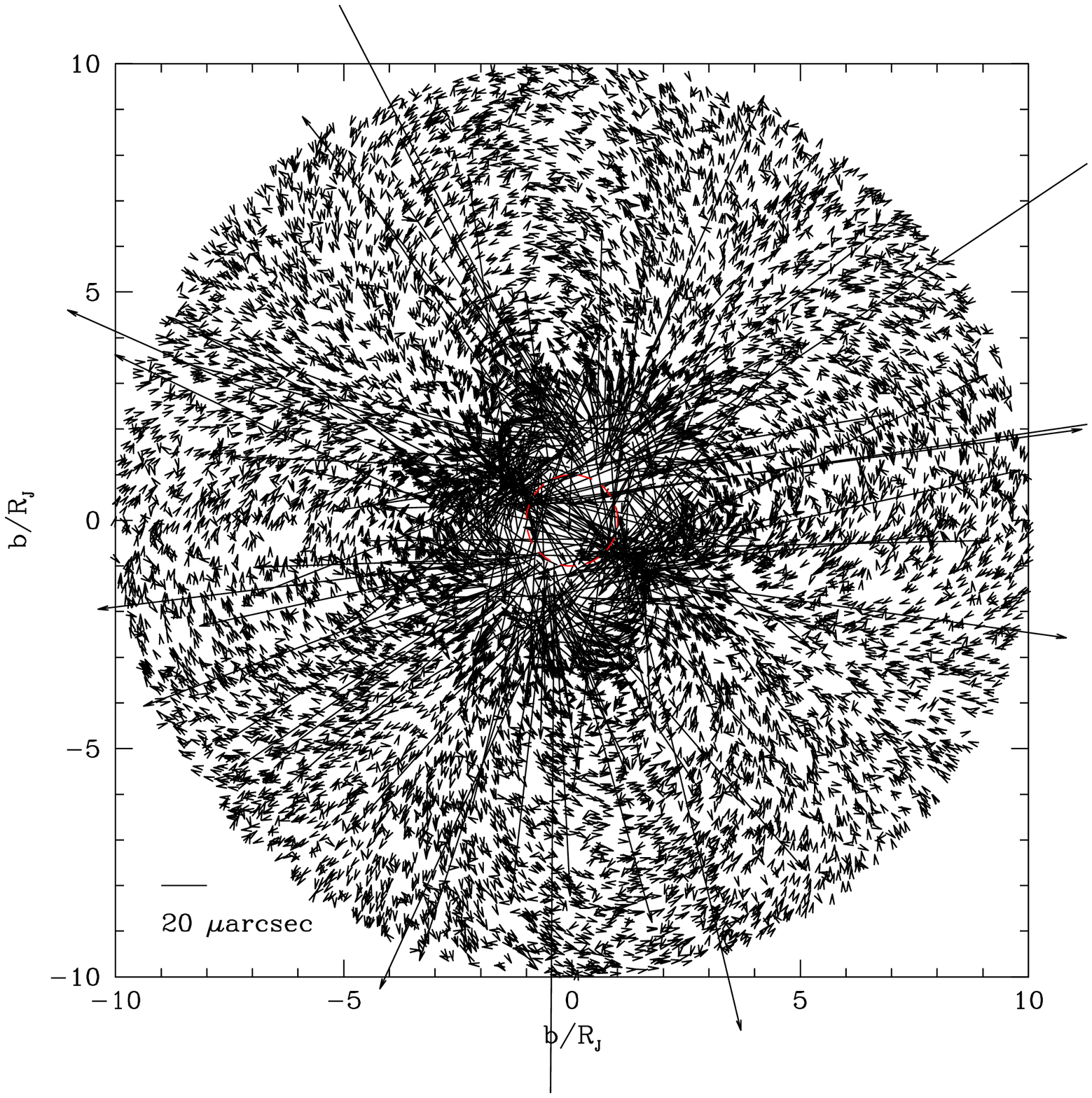}
\caption{
    \label{figure-9-10}The monopole (left-hand side) and
    quadrupole (right-hand side) stellar vector field around Jupiter
    (circle) over the epochs between 2012 and 2018 from the observer
    point of view.  The scales of 10 mas for the monopole and 20
    $\mu$as for the quadrupole are shown on the lower left of the
    plots.} \end{center}
\end{figure}

The plots in~\fref{figure-5-6} show the absolute values of the
monopole and of the quadrupole deflections, obtained with the
simulator for the different epochs and  impact parameters.
\Fref{figure-7-8} is an example of the observed field
corresponding to mid 2013 when Jupiter crosses the Galactic plane.
Even if this date represents a favorable observation, the
relatively limited number of stars close to Jupiter does not lead
to an obvious signature of the stellar displacements produced, in
particular, by the quadrupole effect.  For this reason we have
also plotted the cumulative effect in~\fref{figure-9-10} by
combining all the epochs.  As expected one sees clearly the radial
nature of the monopole deflection and the more complex pattern of
the quadrupole effect.

\section{Measurement of the deflection}
\label{section-4}

\subsection{Processing the observations for $\gamma$ and $J_2$}
\label{section-4.1}

The observed quantities used to fit the relativistic effect in the
light deflection are the along-scan positional differences of the
stars referred to a common origin, namely the center of gravity of
Jupiter. More precisely, we have estimated the measurements
($\Delta \mathbf{\Phi}_{als}$) along  scan  between two transits:
the first one with Jupiter ($\mathbf{\Phi}_{+J}$) and the second
one without Jupiter ($\mathbf{\Phi}_{-J} $), \textit{i.e.}:
\begin{equation}
\Delta \mathbf{\Phi}_{als}= \mathbf{\Phi}_{+J}-\mathbf{\Phi}_{-J}.
\label{eq-obs}
\end{equation}
As stated by~\eref{eq-obs}, the method rests only on the comparison
between small fields taken at two distinct epochs (each point on the
sky will be mapped at least three times during six months) and it is
largely independent from the attitude reconstruction of Gaia. This
makes possible, in estimating the final accuracy, to consider only
random errors and not the systematic attitude errors shared by all the
sources at the two epochs.

The set of $n$ observations can be considered as a system of $n$
equations where the unknowns are the parameters $\gamma$ and
$\epsilon$.  The condition matrix is calculated by evaluating the
partial derivatives with respect to $\gamma$ and $\epsilon$
in~\eref{eq-obs}. The large number of stellar measurements, about
160,000 for 90 Jupiter observations scattered over the 5-year mission
is solved for the two unknowns with a weighted least squares procedure
using singular value decomposition.  The off-diagonal terms of the
final covariance matrix are not significant, indicating a very weak
correlation between the two fitted parameters. Finally anomalous
observations are filtered out with an iterative Student ratio test on
the residuals.

\begin{figure}[htp]
\includegraphics[scale=0.4]{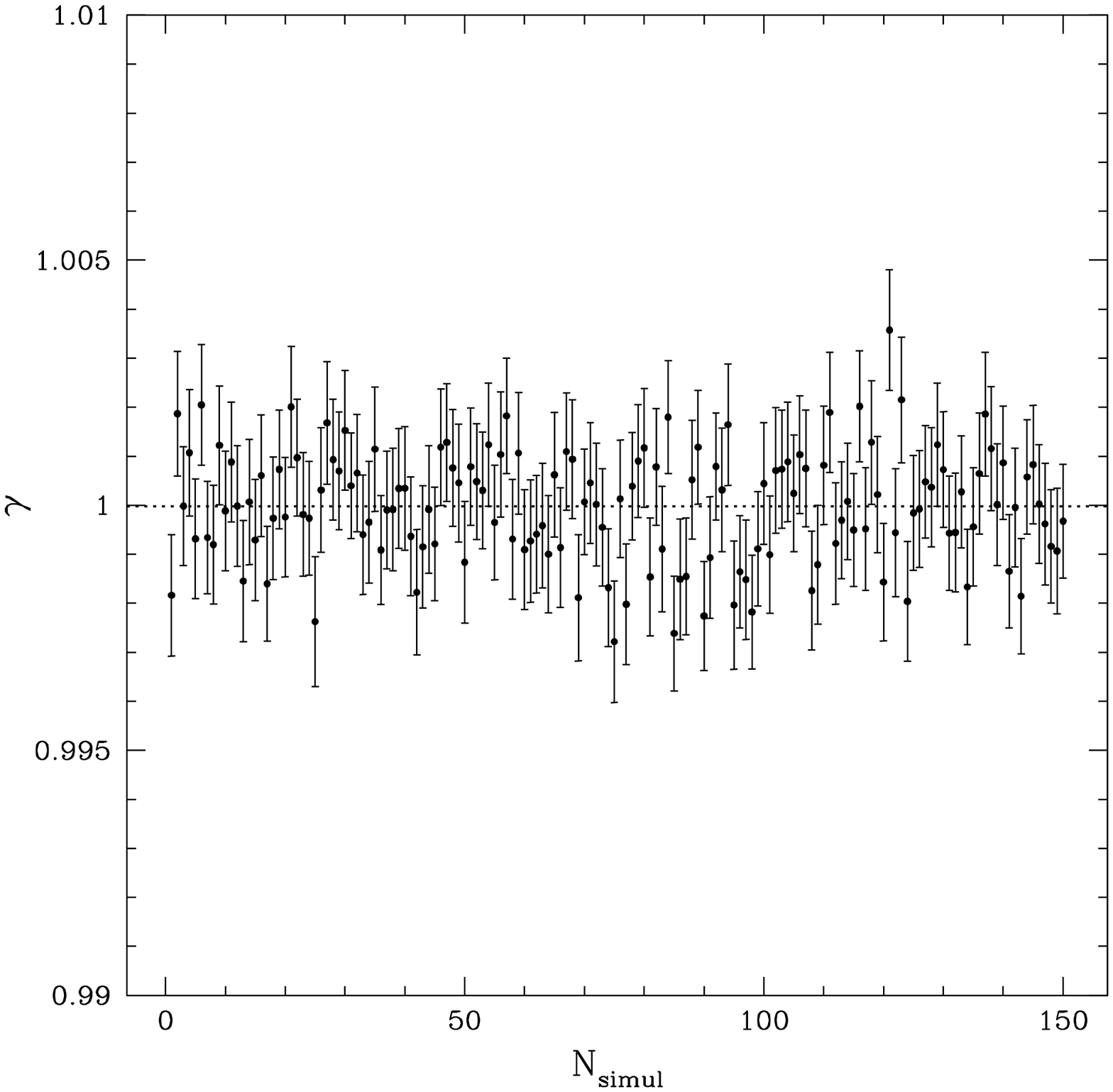}\includegraphics[scale=0.4]{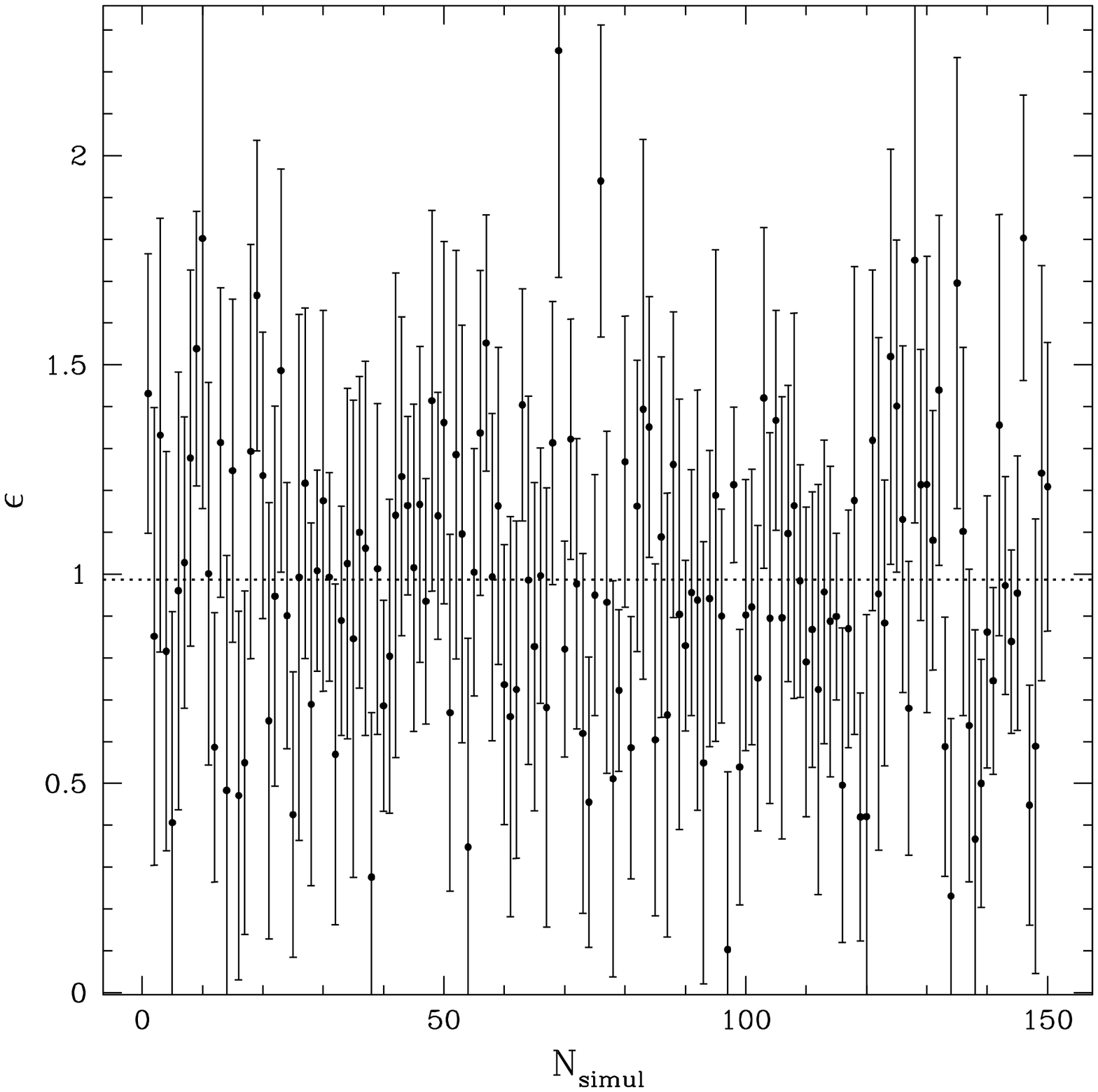}
\begin{center}
\caption{ \label{figure-11-12}The distribution of the mean values of
  $\gamma$ (left-hand side) and $\epsilon$ (right-hand side)
over 150 Montecarlo simulations. The bars are the standard deviations.}
\end{center}
\end{figure}

The estimate of the accuracy with which Gaia could determine
$\gamma$ and $\epsilon$ from small field astrometry has been
estimated by running more than 100 times this numerical simulation
with random drawings stochastically independent in each
experiment. The average of the~100 values of $\gamma$ and
$\epsilon$ prove that there is no bias in the determination as
seen in \fref{figure-11-12}. As for the precision, the scatter was
taken as a more robust way to assess the accuracy than the formal
error of the diagonal terms of the covariance matrix. The mean and
scatter have been evaluated by using the usual point estimates
valid for $n_{run} \gg 1$:

\begin{eqnarray}
<\gamma> &=& \frac{1}{n_{run}} \sum_{i=1}^{n_{run}} \gamma(i) \label{gamma}\\
<\epsilon>&=& \frac{1}{n_{run}} \sum_{i=1}^{n_{run}} \epsilon(i) \label{epsilon}\\
\sigma_{\gamma}&=& \sqrt{\frac{1}{n_{run}} \sum_{i=1}^{n_{run}}
\gamma(i)^2 - {<\gamma>}^2} \label{sigma-gamma} \\
\sigma_{\epsilon}&=& \sqrt{\frac{1} {n_{run}} \sum_{i=1}^{n_{run}}
\epsilon(i)^2 - {<\epsilon>}^2}. \label{sigma-epsilon}
\end{eqnarray}

\subsection{Results}
\label{section-4.2}
The main results are listed in~\tref{table-3} with the
average and scatter on $\gamma$ and $\epsilon$ for three independent
Montecarlo runs.
\begin{table}[htp]
  \caption{\label{table-3} Results for the mean value and
  scatter of ${\gamma}$ and $\epsilon$ in Montecarlo experiments of different lengths.}  \leavevmode
  \begin{center}
       \begin{tabular}{ccccc}
      \hline \\ [-5pt]
       $<\gamma>$ & $ \sigma_{\gamma}$ & $<\epsilon      >$ &
       $\sigma_{\epsilon      }$  & runs \\
\hline\\ [-5pt]
0.9910 &  1.0295$\times 10^{-3}$  & 1.012  & 0.335 & 50 \\
1.0000 &  1.1473$\times 10^{-3}$  & 0.992  & 0.358 & 100 \\
1.0000 &  1.1456$\times 10^{-3}$  & 0.986  & 0.361 & 150 \\
\hline
   \end{tabular}
  \end{center}
\end{table}

It is found with all the assumptions described in the previous
sections that  $\gamma$  can be determined from the monopole light
deflection by Jupiter with an uncertainty of $1.1\times 10^{-3}$.
This is three times better than the optical determination achieved
by Hipparcos, the previous astrometric ESA mission, with stellar
astrometry at large angles from the Sun~\cite{froesch} and 5
million observations of stars. The interest rests not only  on the
accuracy achieved (the global astrometry should reach four orders
of magnitude better  with solar light-bending), but on the fact
that (i) it can be done with a planet, (ii) it is a prerequisite
to the detection of the quadrupole effect in the residuals, (iii)
it opens the way for testing the accurate modelling of the
deflection by a moving body (it is stationary in our experiments).

The final result in $\epsilon $ confirms that the gravitational
effect due to $J_2$ is detectable with Gaia (albeit marginally),
with typically $\epsilon = 1 \pm 0.35$, that is to say a
3-$\sigma$ detection. The effect of  $J_4$ is negligible since it
is much smaller than the random noise on most stars. Using $J_4$
$\sim -0.0006 \sim 4J_2/100$ this results for a grazing ray into a
deflection of $10\mu$as. In addition the effect of $J_4$ decreases
as $(R_J/r)^5$ instead of $(R_J/r)^3$ and very few stars close to
the planet limb would contribute to the signal.

\subsection{Best strategy for the actual experiment}
\label{section-4.3}

The above results concern a statistical analysis over all the epochs
of observations during five years, it is however useful to see whether
some controlled parameters are more favorable than others to observe
the light bending effect and if one can restrict the risk of
systematic errors. First of all we have considered the evolution of
the errors on $\gamma$ and $\epsilon$ with the limiting magnitude, for
various impact parameters and for all the epochs. As we include
fainter stars, they get more numerous but each with a smaller weight
due to the degradation of the astrometric accuracy. It was not obvious
what effect would overcome the other: could many faint stars
contribute significantly or not? The risk being to increase
significantly the chance of systematic deviation for almost no
improvement in the solution.

\begin{figure}[htp]
\begin{center}
\includegraphics[scale=0.4]{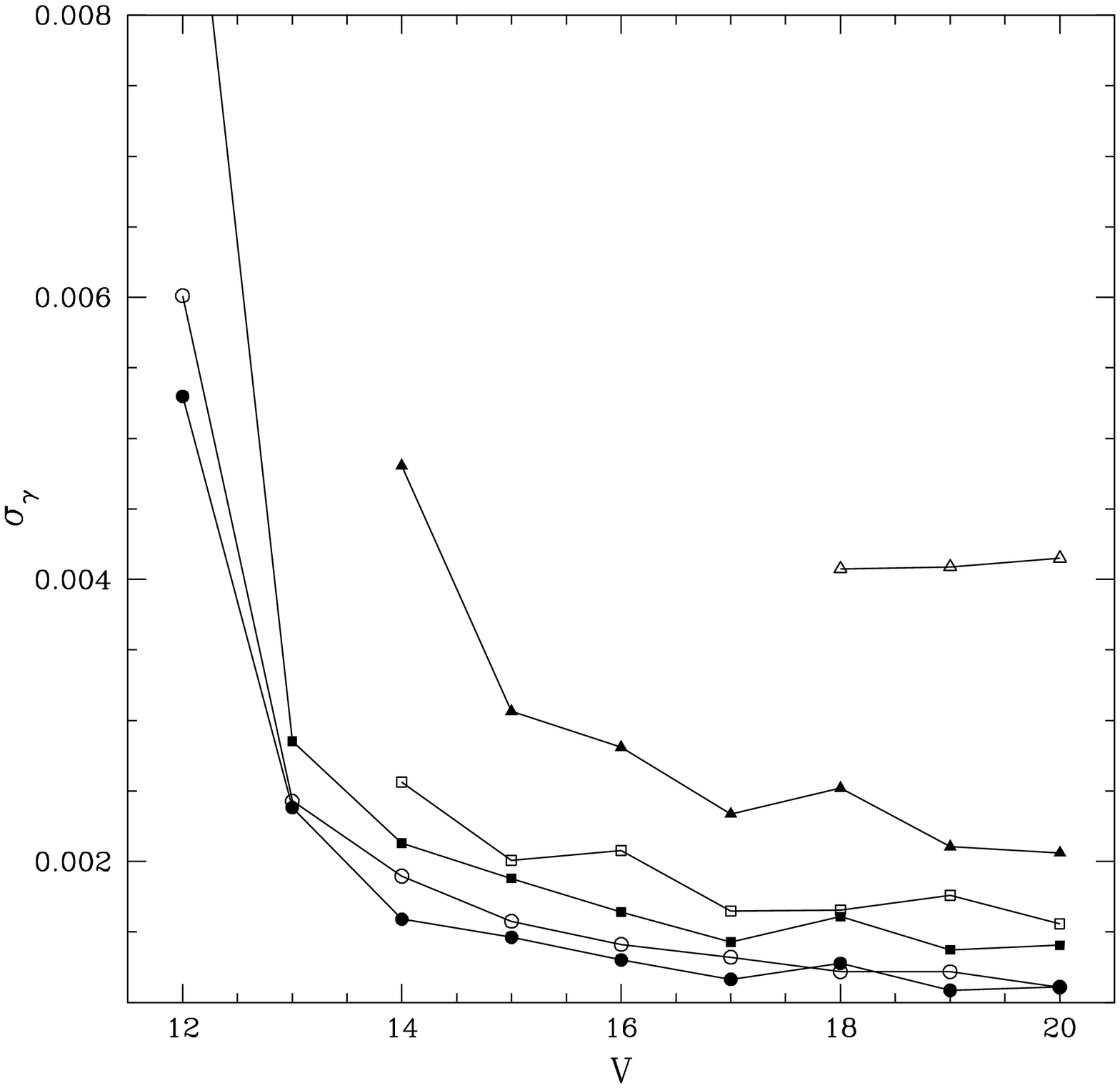}\includegraphics[scale=0.4]{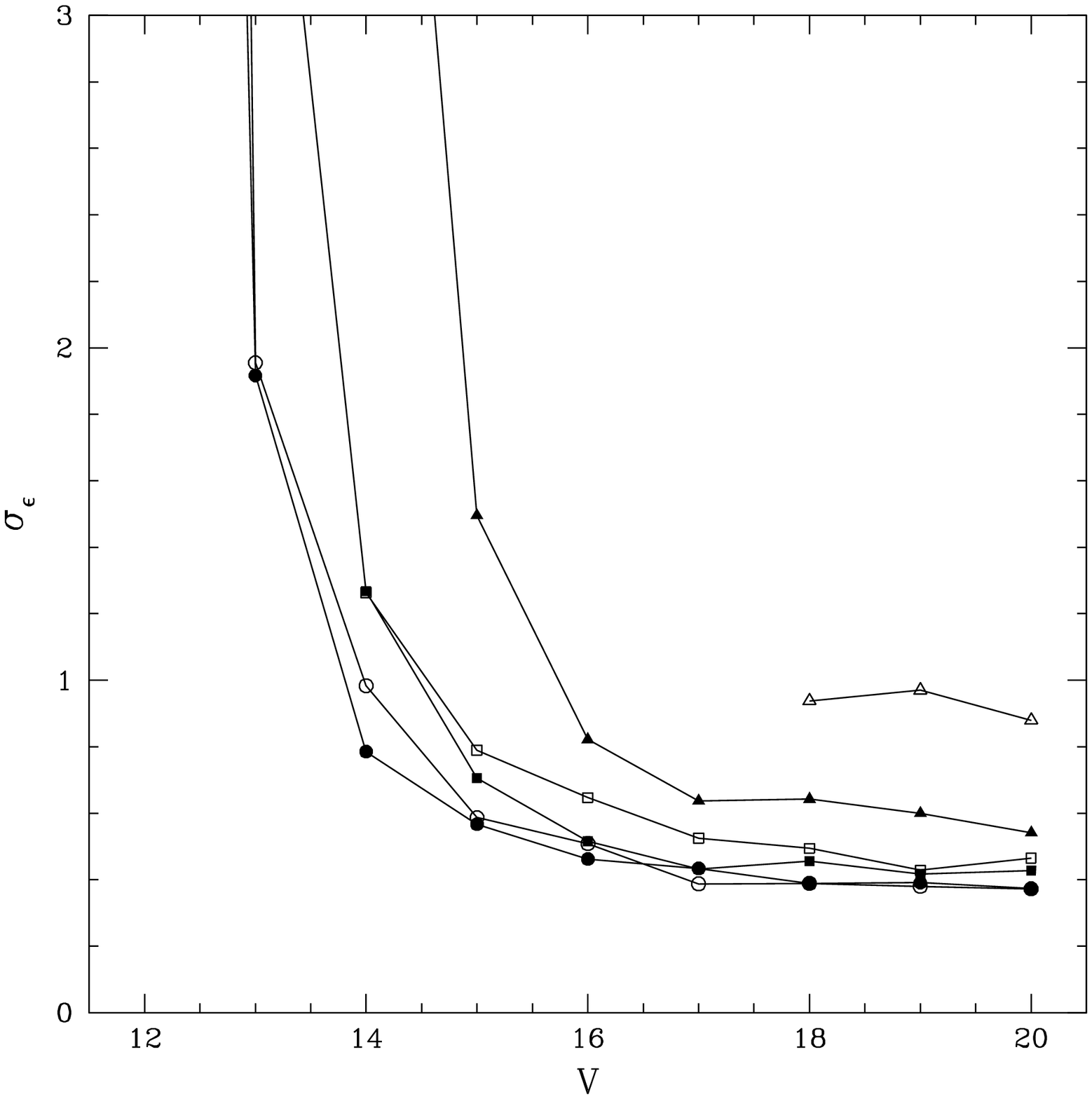}
 \caption{ \label{figure-13-14} Evolution of the errors on $\gamma$
and on $\epsilon$ with the magnitude and for various impact
parameters. Full circles indicate the complete field corresponding
to 45 $R_J$, open circles 30 $R_J$, full squares 20 $R_J$, open
squares 15 $R_J$, full triangle 10 $R_J$ and, open triangle 5
$R_J$.}
\end{center}
\end{figure}

\begin{figure}[htp]
\begin{center}
\includegraphics[scale=0.4]{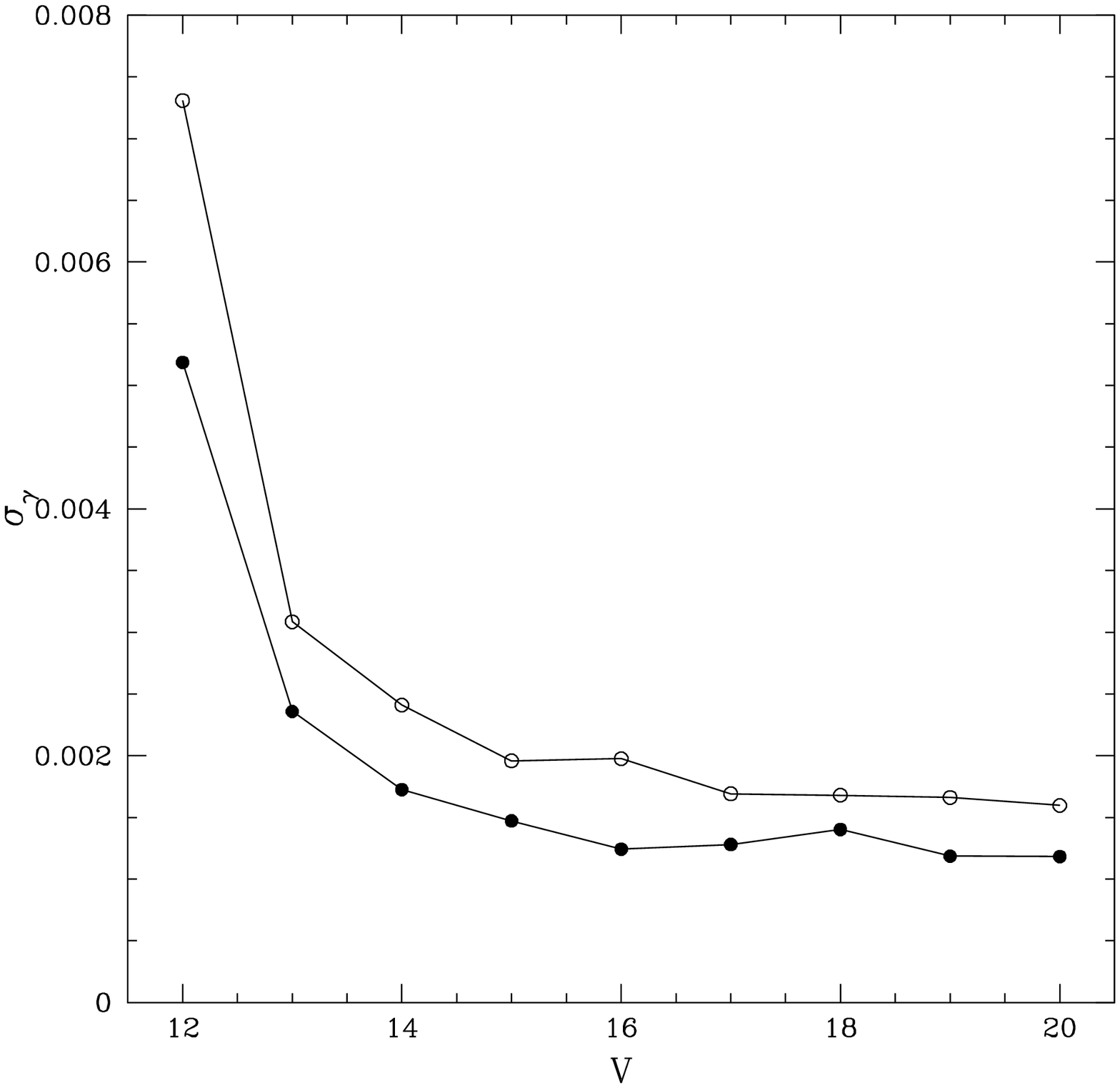}\includegraphics[scale=0.4]{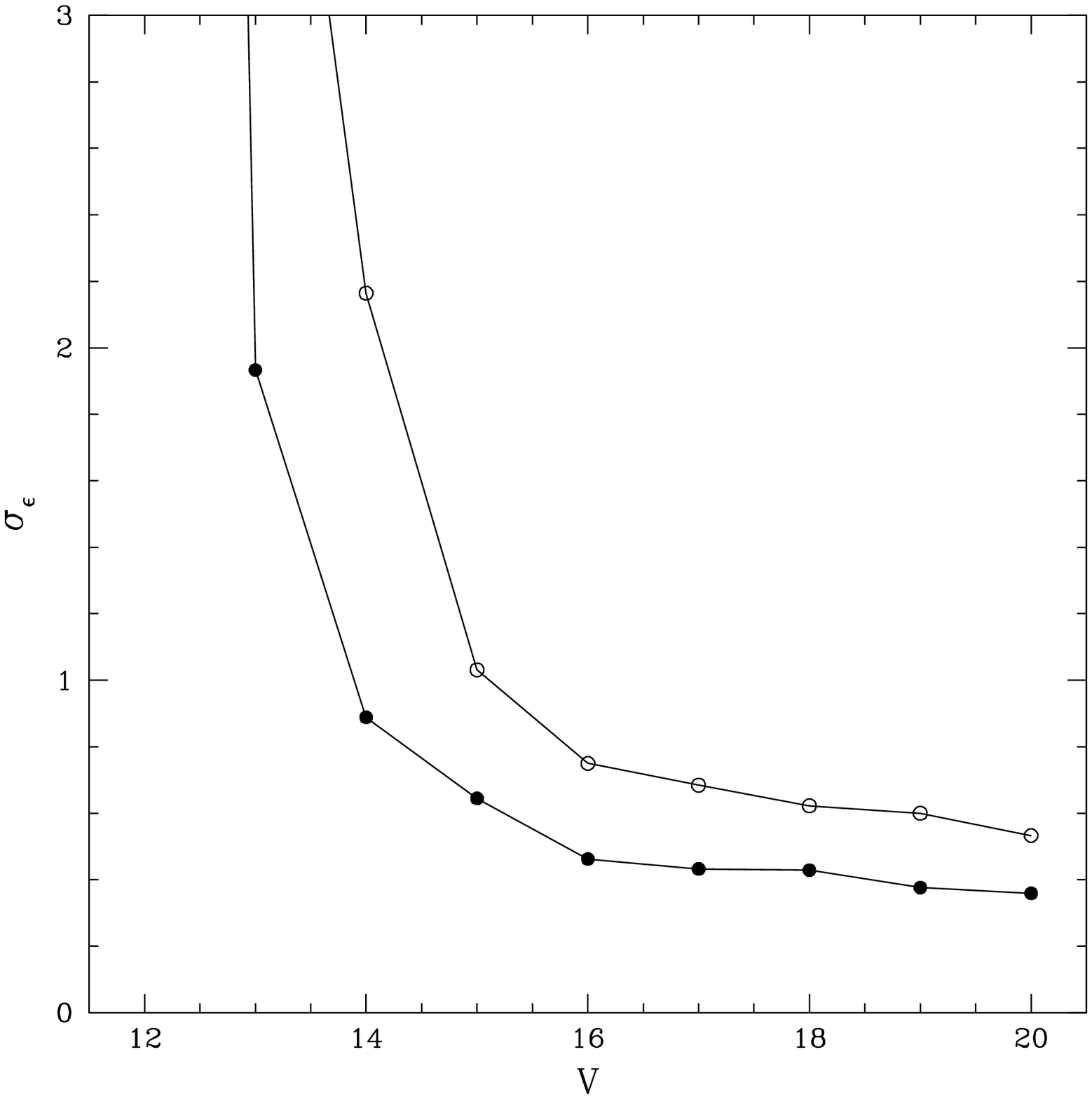}
 \caption{ \label{figure-15-16}The evolution of the errors with
magnitude, keeping the whole Gaia field for all epochs (full circles)
and for 2013 (open circles), when we have the maximum number of
stars. }\end{center}
\end{figure}

The results are plotted in ~\fref{figure-13-14}, where the
precision is shown when only stars brighter than $V$ are used in
the fit. Each curve corresponds to a limitation in the size of the
field of view around Jupiter. It's clear (and not surprising) that
using the full Gaia field provides the best results, in particular
for $\gamma$ just because the monopole deflection decreases more
slowly with increasing impact parameter than the quadrupole's.
However going further than $30 R_J$ brings very marginal
improvement and this can be used to determine what a small field
experiment means in practice. Conversely below 10 $R_J$ the number
of stars decreases very quickly and at 5 $R_J$ we usually have no
star brighter than V=17.5.  As for the magnitude it seems that one
could also reject stars (in case of calibration problems for
example) fainter than $V=16$ without much loss as the final
precision on $\gamma$ and $\epsilon$ does not improve very much
when they are included. That's a very nice feature when it comes
the time to decide on whether faint stars are useful or not. The
graphs show also a degradation in precision when only the
brightest stars are considered, probably because the number of
stars becomes too small to be significant and
formula~\eref{accuracy} inserts a constant error in this case.
Then another interesting experiment has been run to test what
would be the result if one selected only the few epochs during
which we have the maximum number of stars in the background field,
that is to say around 2013 when Jupiter is close to the Galactic
plane. To this purpose, ~\fref{figure-15-16}  shows again the
evolution of the errors with magnitude, when all observations at
every epoch are included (full circles) or only during the
crossing of the Galactic plane in 2013-2014 (open circles with
only 20 epochs instead of 90 in the nominal experiment). Not
surprisingly the errors are larger when considering only the best
epochs, but not dramatically larger, showing that a strategy may
be applied to perform valuable and dedicated experiments during
the first two years of the mission, to validate the concept and
get already significant results.

Finally one should also look at the passage of Jupiter in the middle
of a concentrated open cluster with stars brighter than $V =
13$. There are many such clusters on the ecliptic in the direction of
the Sagittarius, that Jupiter will cross in 2019.  This is after the
nominal mission completion, but well within the possible extended
mission permitted by the consumable. This kind of experiment would
obviously add in the science case to support this extension of the
operations. To this aim we have run the simulator with bright stars
and a surface density comparable to that found in these open
clusters. Results for the determination of $\gamma$ and $\epsilon$ are
given in~\tref{table-4} and~\tref{table-5} with all
the stars of 12 or 13 magnitude respectively. Two realistic
concentrations have been considered around Jupiter corresponding to
fields of 3 and 6 arcmin around the planet.  One sees that a single
experiment can do almost as well as the 5-year mission, provided the
number of bright stars in the cluster is large enough. Note that we
have excluded a priori all the stars with magnitude range between 14
and 16, which statistically contribute significantly to the
precision. At this point this is just an indication showing that a
detail investigation on specific clusters is worth doing.

\begin{table}[htp]
  \caption{\label{table-4}Precisions computed after
  150 Montecarlo runs selecting special fields in 2013
  and increasing artificially the number of stars (n*) (V=12).}
  \leavevmode
  \begin{center}
       \begin{tabular}{lrccc}
      \hline \\  [-10pt]
    &  n* &$ \sigma_{\gamma}$ &  $\sigma_{\epsilon}$  \\
\hline \\ [-10pt]
 $b= 7 R_J$  &   &   &   \\
\hline\\ [-10pt]
  & 1  & 7.41 $\times 10^{-3}$  & 14.54  \\
  & 5 & 3.37 $\times 10^{-3}$  & 2.73 \\
  & 13 & 1.62 $\times 10^{-3}$  & 0.69  \\
  & 20 & 1.39 $\times 10^{-3}$  & 0.34  \\
  & 27 & 1.06 $\times 10^{-3}$  & 0.32  \\
  & 34 & 9.31 $\times 10^{-4}$  & 0.27  \\
\hline \\ [-10pt]
 $b= 15 R_J$  &   &   &   \\
\hline\\ [-10pt]
  & 7  & 4.48 $\times 10^{-3}$  & 9.09  \\
  & 23 &  2.12 $\times 10^{-3}$  &  1.51 \\
  & 39 &  1.70$\times 10^{-3}$  & 0.94  \\
  & 63 & 1.31 $\times 10^{-3}$  & 0.62  \\
  & 79 & 1.26 $\times 10^{-3}$  & 0.51  \\
  & 119 & 9.16 $\times 10^{-4}$  & 0.30  \\
\hline
   \end{tabular}
  \end{center}
\end{table}

\begin{table}[htp]
  \caption{\label{table-5}Precisions computed after
  150 Montecarlo runs selecting special fields in 2013 and increasing
  artificially the number of stars (n*) (V=13).}  \leavevmode
  \begin{center}
  \begin{tabular}{lrccc}
\hline \\ [-10pt]
  &  n* &$ \sigma_{\gamma}$ &  $\sigma_{\epsilon}$  \\
\hline \\ [-10pt]
 $b= 7 R_J$  &   &   &   \\
\hline\\ [-10pt]
  & 9  & 2.49 $\times 10^{-3}$  & 1.51  \\
  & 29 & 1.16 $\times 10^{-3}$  & 0.37 \\
  & 49 & 9.53 $\times 10^{-4}$  & 0.26  \\
  & 66 & 7.50 $\times 10^{-4}$  & 0.20  \\
  & 91 & 6.50 $\times 10^{-4}$  & 0.15  \\
\hline \\  [-10pt]
 $b= 15 R_J$  &   &   &   \\
\hline\\ [-10pt]
  & 44  & 1.88 $\times 10^{-3}$  & 1.28  \\
  & 67  &  1.56 $\times 10^{-3}$  & 0.62 \\
  & 81  &  1.43 $\times 10^{-3}$ & 0.58  \\
  & 97 & 1.17 $\times 10^{-3}$  & 0.41  \\
  & 127 & 1.09 $\times 10^{-3}$  & 0.35  \\
 \hline
     \end{tabular}
  \end{center}
\end{table}

\section{Conclusion}
\label{section-5}
The objectives of this paper has been to study a new way to determine
the PPN parameter $\gamma$ by exploiting a method of differential
positional measurements around Jupiter and to assess the detection of
the light bending from the quadrupole moment of Jupiter.  Thanks to
the high astrometric accuracy to be achieved by the ESA astrometry
mission Gaia and the repeated observations over five years this will
be feasible as the new design of Gaia's optical instrument allows to
process stellar observations very close to the surface of giant
bodies.  We have shown that the deviation to GR with the monopole
deflection can be assessed to $10^{-3}$ with Jupiter only, a result in
the visible better than the optical accuracy already achieved by
Hipparcos with the stars and the solar light bending. With the same
observations, the quadrupole light deflection will be detectable for
the first time with a 3-$\sigma$ confidence level.\\ Although we have
designed an ideal and simplified experiment it includes realistic
observations of Jupiter feasible with Gaia, taking into account the
astrometric accuracy with the star's magnitude deduced from the
current error budget analysis. Whereas the Gaia concept and its design
rest entirely on the global astrometry we have given for the first
time a realistic figure on the true strength of Gaia to carry out also
relativity testing with small field astrometry and propose some
strategies to carry out this experiment in the best conditions. When
the initial conditions of the Gaia scanning law are known, the same
principles will be applied using a real stellar distribution. The
method will be also extended to the observations of Saturn, at least
for the monopole effect.

\ack MTC acknowledges financial support from the Henri Poincar{\'e}
Fellowship during her stay at the Observatoire de la C{\^o}te d'Azur
(Cassiop{\'e}e Department). We wish to thank the members of the
\textit{Relativistic and Reference Frame Working Group} of Gaia
for their valuable comments and collaboration.

\appendix
\section*{Appendix A }
\setcounter{section}{1} This appendix shows the set of fundamental
integrals computed to obtain expressions~\eref{defqu1-chi}
and~\eref{defqu2-chi} for the deflection vector $\Delta
\mathbf{\Phi}$ after having scaled the photon length by the impact
parameter:
\begin{eqnarray}
\int \frac{d \lambda}{(1+ \lambda^2)^{3/2}} &=& \frac{\lambda}{(1+
\lambda^2)^{1/2}}\nonumber \\
\int \frac{d \lambda}{(1+ \lambda^2)^{5/2}} &=& \frac{\lambda/3}{(1+
\lambda^2)^{3/2}}+ \frac{2\lambda/3}{(1+\lambda^2)^{1/2}} \nonumber\\
\int \frac{d\lambda}{(1+ \lambda^2)^{7/2}} &=&
\frac{\lambda/5}{(1+\lambda^2)^{5/2}}+
\frac{4\lambda/15}{(1+\lambda^2)^{3/2}}
+\frac{8\lambda/15}{(1+\lambda^2)^{1/2}}\nonumber \\
\int \frac{\lambda d \lambda}{(1+ \lambda^2)^{5/2}} &=& -\frac{1/3}{(1+
\lambda^2)^{3/2}}\nonumber \\
\int \frac{\lambda d \lambda}{(1+ \lambda^2)^{7/2}} &=& -\frac{1/5}{(1+
\lambda^2)^{5/2}}\nonumber \\
\int \frac{\lambda^2 d \lambda}{(1+ \lambda^2)^{7/2}} &=&-\frac{\lambda/5}{(1+
\lambda^2)^{5/2}}+ \frac{\lambda/15}{(1+\lambda^2)^{3/2}}+
\frac{2\lambda/15}{(1+\lambda^2)^{1/2}} \nonumber
\end{eqnarray}

\section*{Appendix B}

\setcounter{section}{2} Let us consider the angle $\varphi , \vartheta$ of
the Jupiter axis with respect the orthonormal  directions $\mathbf{n}$,
and $\mathbf{t}$, respectively.  From the left hand-side of
equation~\eref{defqu1} we get
\begin{equation}\label{B-mod1}
1 - 2 (\mathbf{n}\cdot \mathbf{z})^2 - 2(\mathbf{t}\cdot\mathbf{z})^2
= - \sin^2
\vartheta \cdot \cos 2 \varphi,
\end{equation}
and from that one of equation~\eref{defqu2}
\begin{equation}\label{B-mod2}
2 (\mathbf{n}\cdot\mathbf{z})(\mathbf{m} \cdot\mathbf{z})=
 \sin^2 \vartheta \sin2\varphi.
\end{equation}
Then, the modulo of the quadrupole deflection term, let's say $
\Delta\mathbf{\Phi} _{quad}$,  in equation~\eref{deftot}) will depend only on $ \sin^2 \vartheta$, namely
on the angle between the $\mathbf{z}$ axis of Jupiter and the direction from
the star to the observer
\begin{equation}
\label{B-quad-mod}
|\Delta \mathbf{\Phi}_{quad}| = \frac{2(1+\gamma) M J_2 R^{3} } {b^3}
\sin^2 \vartheta
\end{equation}
Clearly,  when the $\mathbf{z}$ lies on the \{$\mathbf{n,m}$\} plane,
i.e. $\mathbf{z}$ perpendicular to the line-of-sight, the
quadrupole term will give its maximum contribution.

For a star in direction $\alpha$ with respect to the orthonormal
axis \{$\mathbf{m,n,t}$\} the modulo has a fixed value depending only on the
star's direction of arriving with respect to the above frame
centered on Jupiter, namely:

\begin{equation}
\label{B-quad-n}
|\Delta \Phi_{quad} \mathbf{n} | = \mid \Delta \mathbf{\Phi}_{quad}
\mid \cos2\alpha
\end{equation}
\begin{equation}\label{B-quad-m}
|\Delta \Phi_{quad} \mathbf{m}| = \mid \Delta \mathbf{\Phi}_{quad}
\mid \sin2\alpha.
\end{equation}

\end{document}